\documentstyle[epsfig]{aipproc}

\begin{document}
\title{GEST}

\author{Sun Hong Rhie}
\address{Physics department, University of Notre Dame, In 46556}

\maketitle


\begin{abstract}
Galactic Exoplanet Survey Telescope (GEST) was proposed for a discovery mission 
to search for microlensing terrestrial planets toward the Galactic bulge  
and also Kuiper Belt Objects (KBOs) that are believed to hold vital information 
of the early history of the solar system. Microlensing planet search method is
hinged on photometric singular behavior of lensing (refraction) that is due to
discontinuities in the distribution of photon paths whose phenomenon is better
known in total reflection. The singularities (caustics) directly translate into
potentially large planetary signals but the small size of the caustics imposes 
four essential requirements for earth mass planet searches: massive survey, 
angular resolution, temporal resolution, and continuous monitoring capability 
in a statistically stable observing condition. A 2m scale wide FOV space telescope 
with a large focal plane such as the GEST meets the needs.  

Habitability of earth mass planets in the habitable (liquid water) zone
is suspected to depend crucially on a giant planet near the snowline.
The large mass of the giant planets makes the planetary caustics larger and 
easier to detect them in abundance. In microlensing, one needs not wait 12 years 
to detect a jupiter at the Jupiter orbit (5 AU) unlike in other planet search 
methods (doppler, astrometry, transit) because the planetary microlensing 
events will complete their courses within 50 days or so -- 70 times shorter! 

Close-in giant planets of the bulge stars will be registered as transits in the 
microlensing data base and will form a valuable platform for comparison studies
of the planets from different detection methods and different environments.

When the Galactic bulge is behind the sun, the GEST will be ideal 
for high resolution mapping of the large scale structures which will include 
weak lensing by dark matter, quasar lensing and host galaxies, strong lensing 
by clusters, and a large volume of galaxies, which with selective follow-up 
measurements of whose redshifts and IR images will shed light on the dark stuff 
(matter, energy, essence, extra dimensions, topological defects, ... ).    
\end{abstract}

{\bf Gravitation, Relativities, Microlensing and Planets} \ \  
Microlensing involves both special and general 
relativities in a most trivial way: the spacial warp
due to lensing masses is scanned in time by specical relativistic particles
(photons) emitted from the lensed source and the general relativistic effect
is contained in a factor 2. In microlensing, none of the 3 D's of lensing
-- delay, deflection, and distortion -- are measured unlike in cosmological
lensing because the time of flight differcences and the image sizes
and separations are too small. What is measured is the total magnification of 
the images that varies in time because of the relative motion of the lens and 
the lensed as seen from the observer. Microlensing light curves are essentially 
smooth -- characteristically smooth -- except at caustic crossing discontinuities.
Even the discontinuities are characteristically orderly and smoothed over the size 
of the source (lensed) star. So, given sufficient quality of data, microlensing 
light curves are easily distinguished and identified against the background of 
variable stars and flare stars. 

\begin{figure}[b!] 
\centerline{\epsfig{file=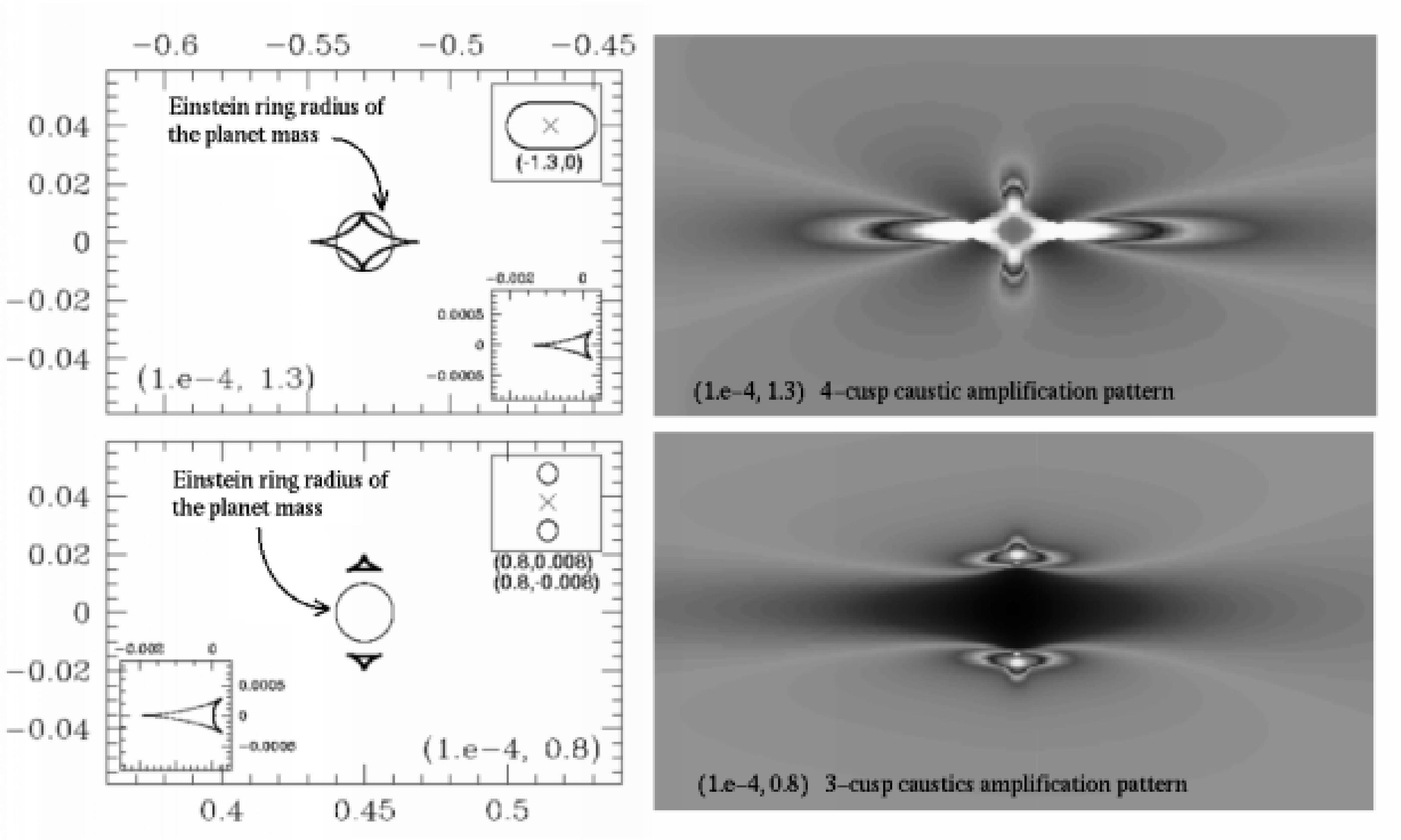,height=3.57in,width=6in}}
\vspace{10pt}
\caption{
A planetary binary lens of a given mass fraction (here $10^{-4}$) has
two types of caustics that contribute to planet detections. When the separation
($\ell$) between the star and the planet is larger than Einstein ring radius
($=1$), the planetary caustic has 4 cusps and is located between the star
and the planet. When $\ell < 1$, two 3-cusped caustics jointly define the
planetary caustic region. Shown are the cases of $\ell =1.3$ and $\ell =0.8$.
A cusp can be considered a point caustic with directionality, and the right
panel exhibits varying magnification power of the cusps. Compare the sizes 
of the caustics and  caustic regions. The Einstein ring of the planet as
a single lens is commonly used as a rough estimator of the planetary caustics 
and is shown at the centers of the caustic planes.    
As a source star traverses the terrain of a planetary caustic, the light 
curve develops telltale ``wavelets". The star is at the coordinate origin and 
the 4-cusped stellar caustics (resembling arrowheads) are shown in the insets. 
The planet positions are shown (marked by $\times$) in relation to the planetary 
critical curves which are centered around  $(-1.3,0)$ and $ (0.8, \pm 0.008)$.
See Bennett and Rhie (1996) for more details of the caustic regions and 
lensing zones.}
\label{texas-fig1}
\end{figure}

A microlensing planetary system will be discovered as a low multiplicity
$n$-point lens. The interference pattern of the Newtonian potentials of 
the masses of the host star(s) and planets determines the angular positions, 
sizes, and strengths of the caustics -- the planetary signal generators. 
The caustic curve of a planetary system lens consists of a stellar caustic 
and planetary caustics. See figure 1 for the case of planetary binary lenses.  
The stellar caustic dominates the behavior of the planetary system as 
a gravitational lens except in the planetary caustic regions.  
The planetary caustics behave like small magnifying glasses that modify 
the would-be single lens light curve by the host star.  The ``size of the 
planetary caustic" (circles in figure 1) of an earth mass planet with mass 
fraction $\epsilon = 3\times 10^{-6}$ is $\approx$ a few $\mu$as.

\begin{figure}[b!] 
\centerline{\epsfig{file=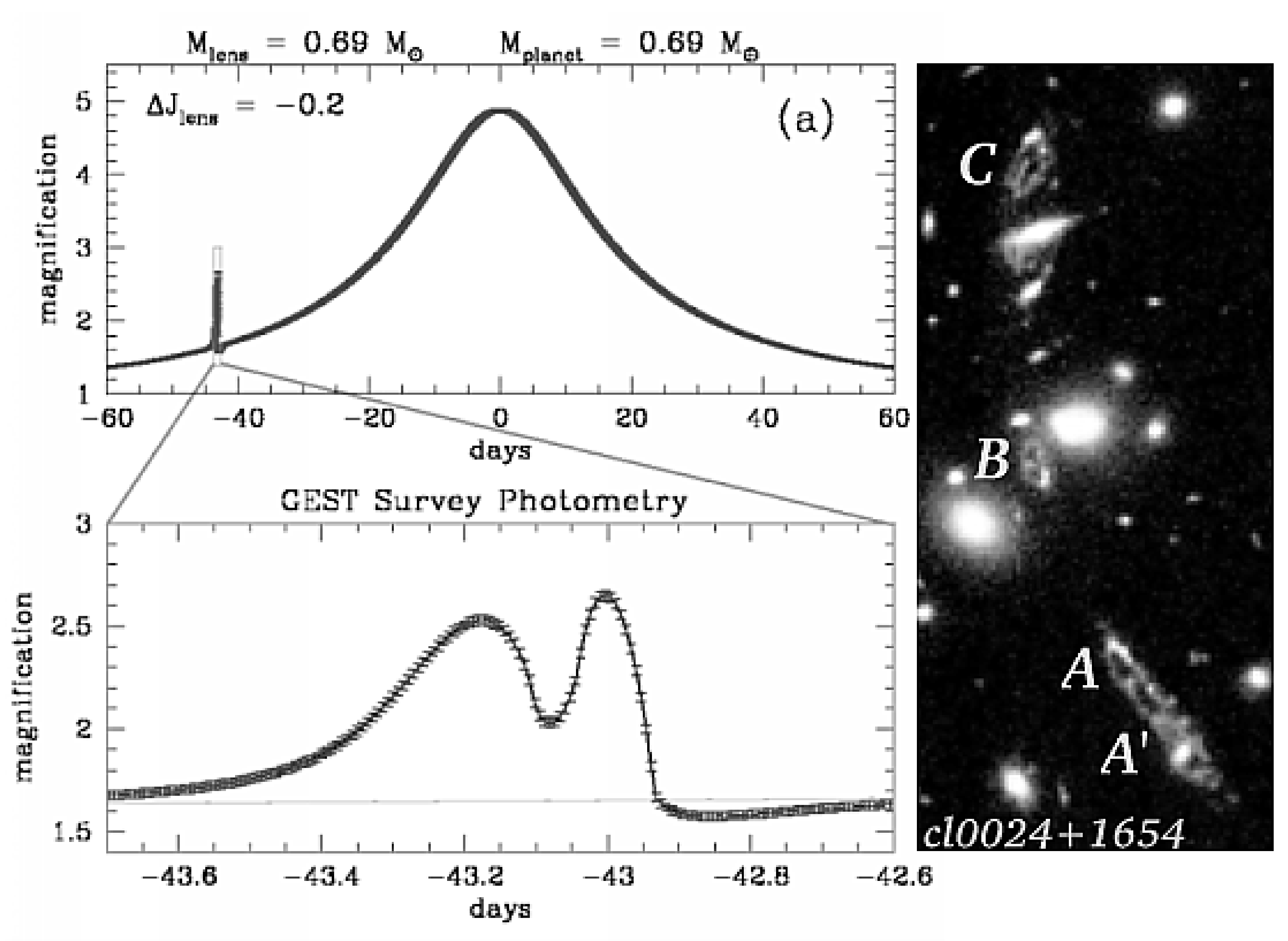,height=4in,width=5.6in}}
\vspace{9pt}
\caption{ The photometrically singular nature of the caustics allows detection 
 of earth mass planets ($\epsilon = 3 \times 10^{-6}$) with large S/N via 
 microlensing. It is essential that the backlighting beam size (size of the 
 source star) is sufficiently smaller than the size of the caustics in order
 to be able to reconstruct the magnification pattern and so the lens parameters
 from light curves. Here the lensed star is a main sequence star and the 
 light curve has been sampled every 10 minutes (Bennett and Rhie 2000).      
 The caustic singularity is associated with creation of two highly magnified 
 images. The blue galaxy images $A$ and $A^\prime$ lensed by cl0024+1654 seem 
 to be one such pair conjoined at the critical curve which we believe is due 
 to local mass concentrations. The image $A^\prime$ was reconstructed in the 
 fit by Tyson et al. (1998) but has so far been ignored by other authors.}
\label{texas-fig2}
\end{figure}
 
\begin{figure}[t!] 
\centerline{\epsfig{file=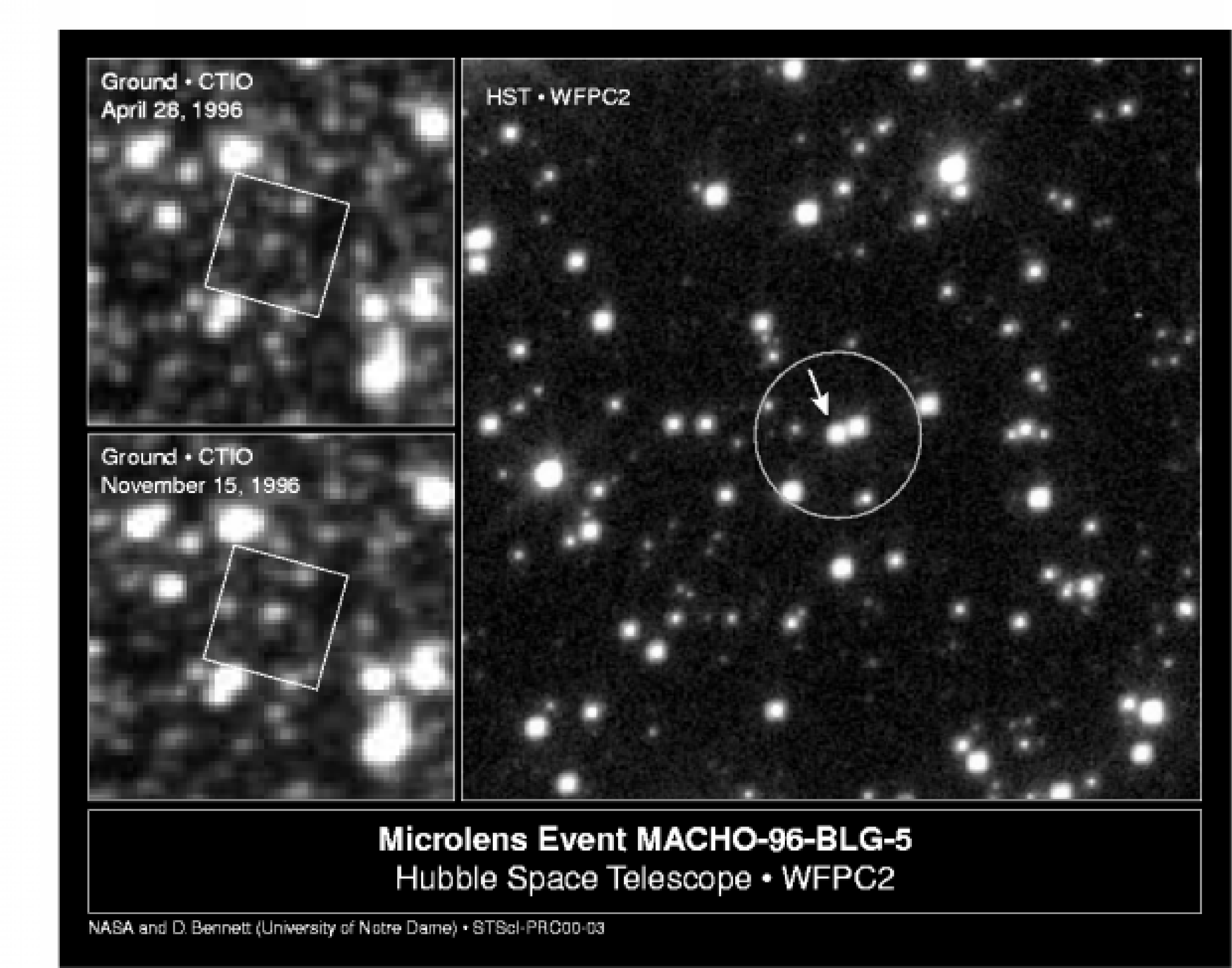,height=3.47in,width=4.3in}}
\vspace{9pt}
\caption{A typical Galactic bulge field is seen from the ground and space.
   The HST frame on the right is the image of the area inside the square 
   boxes on the ground based images (from CTIO) on the left.   
   The circle in the HST image indicates the seeing disk of the CTIO 
   observation shown in the left bottom panel. In the color magnitude
   diagram of gound based observations, the main sequence (MS) stars inside
   the circle in the HST frame will show up as a ``turn-off" star that    
   is a few times brighter than a main sequence star. Since the microlensing 
   beam - size of the Einstein ring radius -- is small $\approx 1$mas, only
   one of the MS stars is lensed (here arrowed) and the blended light
   from the unlensed MS stars adds photon noise and diminishes the effective
   amplification of the observed light curve. For example, a $10\%$ signal 
   will drop to $3.5\%$ with 1.73 times larger error bars.}
\label{texas-fig3}
\end{figure}

\vspace{5pt}
  
  {\bf Ground-based Searches for Earth Mass Planets?} \ \ \ 
  The previous estimations of the feasibility of finding earth mass planets
  from the ground were based on the idea that turn-off stars (brighter than 
  main sequence stars and smaller than giant stars) can be surveyed as source 
  stars \cite{tytler,emplanet,gould,sackett}. The caustic regions in figure 1
  were calculated with the size of a turn-off star ($3 R_\odot$). However,
  it is a false assumption that was derived from a color magnitude diagram 
  of ground-based observations of the crowded Galactic bulge fields.
  The alleged turn-off stars are mostly blended main sequence stars (see
   figure 3). More recent HST data show that turn-off stars are indeed rare 
  \cite{holtz} as one would expect from stellar evolutions. This is 
  detrimental to ground-based searches for earth mass planets because of the
  large seeing disks. It is also the case that the moon closely accompanies the 
  bulge during the high bulge season. With a 4m scale dedicated telesope at the 
  best site (such as dedicated VISTA in Paranal), one can detect earth mass planets, 
  but it takes a very long time to acquire sufficient statistics. The details are 
  being calculated and will be reported in the revised version of \cite{gest}.

\vspace{5pt}
{\bf GEST, Planets, KBOs, and Large Scale Structures}:  \ \ \ 
GEST is an ongoing effort to search from space for microlensing planets in 
the Galactic bulge and the disk. The lack of close-in giants in the globular 
cluster 47 Tucanae indicates the possibility of metallicity constraints on 
the formation of planets or density constraints on the survival of the planets
\cite{gilliland}. Galactic bulge is relatively metal rich ($\approx$ solar 
metallitcy in average) and is expected to be rich with terrestrial planets
which may have been habitable when the Galaxy was younger. Gonzalez et al.
\cite{GHZ} predicts massive terrestrial planets in the bugle, and GEST can 
explore the metallicity effect on the terrestrial planet mass limit. 
Ejection of planets is a commonality according to numerical studies, 
and GEST can detect the free-floating planets further constraining
the planet formation and evolution scenarios. Microlensing is the only 
means to find mature free-floating planets. 

Recent studies of habitability of planetary systems find a strong correlation
between habitable terrestrial planets and jovian planets \cite{lunine}, and
this throws a double jeopardy for most of the planet search efforts: small 
mass of terrestrial planets and long orbital period of jovian planets. 
Microlensing is a peculiar method that can handle the both challenges with ease.

In cosmological lensing, information is extracted from the spatial 
intensity distribution of the photons in the images. In order to reconstruct 
the mass distributions and extract the cosmological parameters accurately,
it is crucial to find systems that offer sufficient number of constraints 
which in turn requires wide and deep imaging with an angular resolution       
of a small fraction of an arcsecond. Excellent systems can be followed up by 
the NGST or ground-based adpative instruements for further information.  

\vspace{9pt}
We thank the GEST/Discovery proposal team members, especially the PI
D. Bennett, and many other supporters of large format space imaging.

\def\ref@jnl#1{{\rm#1}}
\def\aj{\ref@jnl{AJ}}
\def\apj{\ref@jnl{ApJ}}
\def\apjl{\ref@jnl{ApJ}}
\def\apjs{\ref@jnl{ApJS}}
\def\aap{\ref@jnl{A\&A}}
\def\aapr{\ref@jnl{A\&A~Rev.}}
\def\aaps{\ref@jnl{A\&AS}}
\def\mnras{\ref@jnl{MNRAS}}
\def\prl{\ref@jnl{Phys.~Rev.~Lett.}}
\def\pasp{\ref@jnl{PASP}}
\def\nat{\ref@jnl{Nature}}
\def\sci{\ref@jnl{Science}}
\def\iauc{\ref@jnl{IAU~Circ.}}
\def\aplett{\ref@jnl{Astrophys.~Lett.}}
\def\annrev{\ref@jnl{Ann.~Rev.~Astron.~and Astroph.}}

\end{document}